\documentclass[conference]{IEEEtran}
\usepackage[utf8]{inputenc}
\usepackage[T1]{fontenc}
\usepackage{url}
\usepackage{ifthen}
\usepackage{cite}
\usepackage[cmex10]{amsmath}
\usepackage{graphicx}
\usepackage{amssymb,bm}
\usepackage{bbm}
\usepackage{amsmath}
\usepackage{amsfonts}
\usepackage{cite}
\usepackage{enumitem}

\allowdisplaybreaks
\usepackage{amsthm}
\newtheorem{theorem}{Theorem}

\usepackage{chngcntr}
\counterwithout{theorem}{section}
\counterwithout{definition}{section}
\counterwithout{lemma}{section}
\counterwithout{remark}{section}
\counterwithout{assumption}{section}
\counterwithout{proposition}{section}
\counterwithout{corollary}{section}
\IEEEoverridecommandlockouts
\begin{document}
\bstctlcite{IEEEexample:BSTcontrol}
\title{An Alternative Achievability Proof for the DoF Region of the MISO BC with Partial CSIT}
\title{DoF Region of the MISO BC with Partial CSIT: Proof by Inductive Fourier-Motzkin Elimination}

 \author{%
   \IEEEauthorblockN{Hamdi~Joudeh and Bruno~Clerckx}
   \IEEEauthorblockA{Department of Electrical and Electronic Engineering, Imperial College London, United Kingdom\\
                     Email: \{hamdi.joudeh10, b.clerckx\}@imperial.ac.uk}
\thanks{This work is partially supported by the U.K. Engineering and Physical Sciences Research Council (EPSRC) under grant EP/N015312/1.}
 }

\maketitle
\begin{abstract}
We provide a fresh perspective on the problem of characterizing the
DoF region of the $K$-user MISO BC with arbitrary levels of partial CSIT.
In a previous achievability proof, Piovano and Clerckx
characterized all faces describing a polyhedral outer bound region,
and then with the aid of mathematical induction, prescribed a scheme based on rate-splitting with flexible assignment of common DoF and power levels to achieve each such face.
We propose an alternative approach in which we deal directly with the region achievable through rate-splitting and employ a
Fourier–Motzkin procedure to eliminate  all auxiliary variables, hence reducing the achievable region to the known outer bound.
A key insight emerging from our proof is that tuning only one power variable, as well as assigning the common DoF, is sufficient to achieve the entire DoF region, as opposed to $K$ power variables previously employed.
\end{abstract}
\section{Introduction}
\label{sec:introduction}
Degrees-of-freedom (DoF) studies for wireless networks seek to characterize the optimal
number of signalling dimensions accessible at each receiver in the
asymptotically high signal to noise ratio (SNR) regime.
While caution must be practiced in translating DoF findings into practical
insights, such findings nevertheless serve as a crude first step along a path of
refinements towards understanding capacity limits.

With an initial focus on scenarios in which channel state information at the transmitters (CSIT)
is perfectly known, DoF studies have shifted in recent years towards incorporating various forms of CSIT imperfections and
uncertainties \cite{Vaze2012,Yang2013,Tandon2013,Rassouli2016,Davoodi2016}.
A canonical setting for such studies is the $K$-user multiple-input-single-output (MISO)
broadcast channel (BC), as its DoF constitutes an outer bound for more intricate settings,
e.g. the interference channel and the X channel.
Moreover, amongst the various models of CSIT imperfections, the partial CSIT model, which captures
a whole range of uncertainty levels ranging from finite-precision to perfect CSIT,
has recently become of particular research interest largely
due to the difficulty of  the corresponding optimality (converse) proofs \cite{Davoodi2016}.
This paper focuses on  the $K$-user MISO BC with arbitrary levels of partial
CSIT, i.e.  the CSIT estimation error for user $k$ is assumed to scale as $\sim \mathrm{SNR}^{-\alpha_{k}}$, where
$\alpha_{k} \in [0,1]$ is a parameter that controls the corresponding CSIT level.

Assuming, without loss of generality, that $\alpha_{1} \geq \alpha_{k}$ for all $k$,
the sum-DoF of the above channel is given by
\begin{equation}
\label{eq:sum_DoF}
d_{\Sigma} = 1 + \alpha_{2} + \cdots + \alpha_{K}.
\end{equation}
This is achieved through a rate-splitting scheme, with a combination of zero-forcing and multicasting signals,
proposed in \cite{Yang2013} for the 2-user setting and generalized to $K$-user settings in \cite{Clerckx2016} and references therein.
The true challenge in characterizing the sum-DoF in \eqref{eq:sum_DoF} is the converse; a conjecture in \cite{Lapidoth2005} that $d_{\Sigma}$ collapses to $1$ under finite precision CSIT, i.e $\alpha_{k} = 0$ for all $k$, remained open for nearly a decade.
This was finally settled by Davoodi and Jafar in \cite{Davoodi2016}, showing that
$d_{\Sigma}$ is bounded above by $1 + \alpha_{2} + \cdots + \alpha_{K}$
using an unconventional combinatorial argument known as the aligned image sets approach.

Once equipped with the converse in \cite{Davoodi2016}, a polyhedral outer bound for the
entire DoF region can be easily constructed by considering each subset of users, and bounding the sum-DoF for such subset
while eliminating remaining users (see \eqref{eq:DoF_region_single_subchannel}
in Section \ref{sec:preliminaries}).
With this in mind, the main challenge in going from characterizing the sum-DoF to characterizing the entire DoF region now
becomes the achievability side of the argument.
The rate-splitting scheme used to achieve the sum-DoF in \eqref{eq:sum_DoF} is,
in general, specified by several design variables for power control and common DoF
assignment.
While such design variables can be optimized to obtain a DoF tuple that maximizes a certain
scalar objective function, e.g.  the sum-DoF \cite{Joudeh2016} or the symmetric-DoF \cite{Joudeh2016a},
the entire achievable DoF region is generally described as the collection of DoF tuples achieved
through all combinations of feasible design variables.
Matching such achievable DoF region, described using a mixture of fixed CSIT parameters and
auxiliary design variables,  to the outer bound, expressed in terms of fixed CSIT parameters only (e.g. see \eqref{eq:sum_DoF}), is
no trivial task in general.

In \cite{Piovano2017}, Piovano and Clerckx showed that the DoF region achieved
through rate-splitting, with flexible power control and common DoF
assignment, is in fact optimal.
This is accomplished by an exhaustive characterization of all faces describing the DoF outer bound obtained from
\cite{Davoodi2016}, and then prescribing tuned rate-splitting strategies that attain all DoF tuples in each such face,
while using induction over $K$.

In this paper, we take an alternative route to proving the result in \cite{Piovano2017}.
Instead of starting from the outer bound region and showing that its constituent faces are achievable,
we start with the rate-splitting achievable region and prove that it is equivalent to the outer bound
region.
This is accomplished by eliminating the power control and common DoF assignment auxiliary  variables from the
representation of the achievable region through a series of reductions followed by an inductive Fourier-Motzkin
elimination procedure.
An interesting insight that emerges from our new proof is that it is sufficient to optimize only a single
power control variable, in addition to the common DoF assignment variables,
to achieve all points of the DoF region, as opposed to $K$ power control variables
as done in \cite{Piovano2017}.
A further discussion on the distinctions between the two approaches is given in Section \ref{sec:Discussion}.

\emph{Notation:}
$a,A$ are scalars, with $A$ often denoting a random variable unless the contrary was obvious,
$\mathbf{a} \triangleq (a_{1},\ldots,a_{k})$ is $k$-tuple of scalars,
and $\mathcal{A}$ is a set.
For any $\mathcal{S} \subseteq \{1,\ldots,k\}$, we use $\mathbf{a}(\mathcal{S})$ to denote $\sum_{i \in \mathcal{S}}a_{i}$.
For any positive integers $k_{1}$ and $k_{2}$ with $k_{1} \leq k_{2}$,
the sets $\{1,\ldots,k_{1}\}$  and $\{k_{1},\ldots,k_{2}\}$ are denoted by $\langle k_{1} \rangle$ and $\langle k_{1} : k_{2} \rangle$, respectively.
For sets $\mathcal{A}$ and $\mathcal{B}$, $\mathcal{A}\setminus \mathcal{B}$ is the set of elements in
$\mathcal{A}$ and not in $\mathcal{B}$.
\section{System Model}
\label{sec:system_model}
We consider a MISO BC comprising a $K$-antenna transmitter and $K$ single-antenna receivers (users), where the index sets for receivers is given by $\mathcal{K} \triangleq \langle K \rangle$.
For transmissions occurring over $n>0$ channel uses, the channel model is given by the following input-output relationship:
\begin{equation}
\label{eq:received signal}
Y_{k}(t)= \sum_{i \in \mathcal{K}} G_{ki}(t) X_{i}(t)+ Z_k(t), \  k \in \mathcal{K}
\end{equation}
For channel use $t$, $Y_{k}(t)$ is the signal observed by receiver $k$, $G_{ki}(t)$ is the fading channel coefficients between transmit antenna $i$ and receiver $k$, $X_{i}(t)$ is the symbol transmitted from antenna $i$,
and $Z_k(t) \sim \mathcal{N}_{\mathbb{C}}(0,1)$ is the zero mean unit variance additive white Gaussian noise (AWGN) at receiver $k$.
All signals and channel coefficients are complex.
The transmitter is subject to the power constraint $\frac{1}{n} \sum_{t=1}^{n} \sum_{k = 1}^{K} | X_{k}(t) |^{2}
\leq P$.
\subsection{Partial CSIT}
\label{subsec:partial_CSIT}
Under partial CSIT, the channel coefficients associated with receiver $k$ over subchannel $m$ are modeled as
\begin{equation}
\label{eq:channel model}
G_{ki}(t) = \hat{G}_{ki}(t) + \sqrt{P^{-\alpha_{k}}}\tilde{G}_{ki}(t), \ i \in \mathcal{K}, t  \in \langle n \rangle
\end{equation}
where $\hat{G}_{ki}(t)$ and $\tilde{G}_{k}(t)$ are the corresponding channel estimate and estimation error terms, respectively, while
$\alpha_{k} \in [0,1]$ is the CSIT level parameter.
We consider non-degenerate channel situations and a non-degenerate channel uncertainty model according to the definition in \cite{Davoodi2016}, where channel variables $\hat{G}_{ki}(t)$ and $\tilde{G}_{ki}(t)$ are subject to the bounded density assumption.
The difference between $\hat{G}_{ki}(t)$ and $\tilde{G}_{ki}(t)$ is that the realizations of the former
are revealed to the transmitter, while the realizations of the latter are not.
Under this model, the parameter $\alpha_{k}$  captures the whole range of knowledge available at the transmitter of receiver $k$'s channel, i.e.
$\alpha_{k}= 0$ corresponds to situations where channel knowledge is absent and
$\alpha_{k} = 1$ corresponds to perfect CSIT, both in a DoF sense.
Without loss of generality, we assume that
 \begin{equation}
\label{eq:CSIT order}
\alpha_{1} \geq \alpha_{2} \geq \cdots \geq \alpha_{K}.
\end{equation}
The tuple of CSIT levels is given by $\bm{\alpha} \triangleq (\alpha_{1},\ldots,\alpha_{K})$.
\subsection{Messages, Rates, Capacity and DoF}
\label{subsec:msgs-rates-capacity-DoF}
Messages $W_1,\ldots, W_K$,
the corresponding achievable rates $R_1(P),\ldots,R_K(P)$ and the capacity region $\mathcal{C}(P)$ are all defined in the standard Shannon theoretic sense.
The DoF tuple $\mathbf{d} \triangleq (d_{1}, \ldots, d_{K})$ is said to be achievable if there exists $(R_1(P),\ldots,R_K(P)) \in \mathcal{C}(P)$ such that $d_k=\lim_{P \to \infty} \frac{R_k(P)}{\log(P)}$ for all $k \in \langle K \rangle$.
The DoF region is denoted by $\mathcal{D}$, and is defined as the closure of all achievable DoF tuples
$\mathbf{d}$.
\section{Preliminaries}
\label{sec:preliminaries}
In this section, we set the stage for the main result of this paper, presented in the following section.
We start by a (re)statement of \cite[Th. 1]{Piovano2017}, in which $\mathcal{D}$ is characterized.
\begin{theorem}
\label{Theorem:DoF_region}
{\normalfont \textbf{\cite[Th. 1]{Piovano2017}}}. For the MISO BC with partial CSIT described in Section \ref{sec:system_model}, the optimal DoF region
is given by
\begin{equation}
\label{eq:DoF_region_single_subchannel}
\mathcal{D} = \Big\{ \mathbf{d} \in \mathbb{R}_{+}^{K} : \mathbf{d}(\mathcal{S}) \leq 1 + \bm{\alpha}\big(\mathcal{S} \setminus \{\min \mathcal{S}\} \big), \ \mathcal{S} \subseteq \mathcal{K}\Big\}.
\end{equation}
\end{theorem}
Due to \eqref{eq:CSIT order}, for any $\mathcal{S} \subseteq \mathcal{K}$, the corresponding inequality in
\eqref{eq:DoF_region_single_subchannel} is equivalent to $\mathbf{d}(\mathcal{S}) \leq 1 + \bm{\alpha}(\mathcal{S} ) - \max_{i \in \mathcal{S}}\{ \alpha_{i}\}$.

The converse proof for Theorem \ref{Theorem:DoF_region}, i.e. showing that the right-hand-side of \eqref{eq:DoF_region_single_subchannel} is an outer bound for the DoF region, is a direct consequence of the result in \cite[Th. 1]{Davoodi2016}.
The contribution of \cite{Piovano2017} is proving that the region in \eqref{eq:DoF_region_single_subchannel}
is in fact achievable.
Next, we present our own take on the approach in \cite{Piovano2017}, focusing on parts most essential for introducing and appreciating
the alternative approach presented in the following section.

The achievability of $\mathcal{D}$ is based on
rate-splitting with the superposition of private (zero-forcing) and common (multicasting) codewords \cite{Clerckx2016}.
The DoF region achieved through this scheme, which we denote by $\mathcal{D}_{\mathrm{RS}}^{\star}$,
is characterized as the set of all DoF tuples $\mathbf{d} = (d_{1},\ldots,d_{K}) \in \mathbb{R}_{+}^{K}$ that satisfy
\begin{subequations}
\label{eq:DoF_region_RS}
\begin{align}
  &(d_{1},\ldots,d_{K}) = (d_{1}^{(\mathrm{p})},\ldots,d_{K}^{(\mathrm{p})}) + (d_{1}^{(\mathrm{c})},\ldots,d_{K}^{(\mathrm{c})}) \\
  & d_{i}^{(\mathrm{p})} \geq 0, \ d_{i}^{(\mathrm{c})} \geq 0, \ i \in \mathcal{K} \\
  &d_{i}^{(\mathrm{p})} \leq \Big( a_{i} -  \big( \max_{j \neq i}  a_{j}  - \alpha_{i}   \big)^{+}
  \Big)^{+}, \ i \in \mathcal{K} \\
  &\sum_{i \in \mathcal{K}} d_{i}^{(\mathrm{c})} \leq 1 - \max_{j \in \mathcal{K}}a_{j} \\
  &0 \leq a_{i} \leq  1, \ i \in \mathcal{K}.
\end{align}
\end{subequations}
In the above, $\mathbf{d}^{(\mathrm{p})} \triangleq (d_{1}^{(\mathrm{p})},\ldots,d_{K}^{(\mathrm{p})}) \in \mathbb{R}_{+}^{K}$  and
$ \mathbf{d}^{(\mathrm{c})} \triangleq (d_{1}^{(\mathrm{c})},\ldots,d_{K}^{(\mathrm{c})}) \in \mathbb{R}_{+}^{K}$
are the private and common DoF tuples, associated with the private and common parts of the $K$ messages,
respectively, which result from message splitting.
On the other hand, $\mathbf{a} \triangleq (a_{1},\ldots,a_{K}) \in [0,1]^{K}$ are the power control variables associated with the $K$ private signals, i.e. the power assigned to the $k$-th private signal scales as $\sim P^{a_{k}}$.
For a detailed exposition of this scheme and its achievable DoF in \eqref{eq:DoF_region_RS}, readers are referred to \cite{Clerckx2016,Piovano2017} and references therein.

From \eqref{eq:DoF_region_RS}, it is evident that each DoF tuple $\mathbf{d}\in\mathcal{D}_{\mathrm{RS}}^{\star}$ is achieved through a strategy
identified by a pair $\big(\mathbf{a},\mathbf{d}^{(\mathrm{c})}\big)$, where the power control tuple $\mathbf{a}$ determines
the private DoF tuple $\mathbf{d}^{(\mathrm{p})}$ and the common sum-DoF $\mathbf{d}^{(\mathrm{c})}(\mathcal{K}) = \sum_{i \in \mathcal{K}} d_{i}^{(\mathrm{c})}$, while the individual entries of $\mathbf{d}^{(\mathrm{c})}$ determine the manner in which the common sum-DoF is assigned across the $K$ users.

As it turns out, the achievable DoF region described in \eqref{eq:DoF_region_RS} is equivalent to the outer bound in \eqref{eq:DoF_region_single_subchannel}, i.e. $\mathbf{d} \in \mathcal{D}_{\mathrm{RS}}^{\star}$ if and only if  $\mathbf{d} \in \mathcal{D}$.
This is shown in \cite{Piovano2017} by explicitly tuning the strategy, i.e. the pair $\big(\mathbf{a},\mathbf{d}^{(\mathrm{c})}\big)$,
to achieve each and every face of the polyhedral outer bound in \eqref{eq:DoF_region_single_subchannel}.
This involves an exhaustive characterization of the faces of the polyhedron $\mathcal{D}$, in which an induction argument is employed
to cope with an arbitrary dimension (i.e. number of users) $K$.
\section{Auxiliary Variable Elimination Approach}
The system of inequalities in \eqref{eq:DoF_region_RS} can be viewed as some region in an extended
$4K$-dimensional space, with coordinates represented by $(\mathbf{d},\mathbf{d}^{(\mathrm{p})},\mathbf{d}^{(\mathrm{c})},\mathbf{a})$.
We essentially wish to project out (or eliminate) the variables
$(\mathbf{d}^{(\mathrm{p})},\mathbf{d}^{(\mathrm{c})},\mathbf{a})$  from \eqref{eq:DoF_region_RS}
and obtain a description of the $K$-dimensional region $\mathcal{D}_{\mathrm{RS}}^{\star}$ in terms of
the variables $\mathbf{d}$ and the fixed parameters $\bm{\alpha}$ only.

We start by applying some simplifying reductions (in the form of restrictions) to \eqref{eq:DoF_region_RS}.
In particular, we restrict the tuple of power control variables  such that
$\mathbf{a} = (a,\ldots,a)$, where $a \in [0,1]$. By doing so, we essentially reduce the $K$ power control variables $(a_{1},\ldots,a_{K})$ in $\mathcal{D}_{\mathrm{RS}}^{\star}$ to a single variable $a$.
The resulting achievable region, denoted by $\mathcal{D}_{\mathrm{RS}}$, is given by all DoF tuples $\mathbf{d} = (d_{1},\ldots,d_{K}) \in \mathbb{R}_{+}^{K}$
that satisfy
\begin{subequations}
\label{eq:DoF_region_RS_a}
\begin{align}
  &(d_{1},\ldots,d_{K}) = (d_{1}^{(\mathrm{p})},\ldots,d_{K}^{(\mathrm{p})}) + (d_{1}^{(\mathrm{c})},\ldots,d_{K}^{(\mathrm{c})}) \\
  & d_{i}^{(\mathrm{p})} \geq 0, \ d_{i}^{(\mathrm{c})} \geq 0, \ i \in \mathcal{K} \\
  \label{eq:DoF_region_RS_a_min}
  &d_{i}^{(\mathrm{p})} \leq \min\{a,\alpha_{i}\}, \ i \in \mathcal{K} \\
  &\sum_{i \in \mathcal{K}} d_{i}^{(\mathrm{c})} \leq 1 - a \\
  &0 \leq a \leq  1.
\end{align}
\end{subequations}
Due to the additional constraints of $a_{i} = a$, $i \in \mathcal{K}$, it is evident that $\mathcal{D}_{\mathrm{RS}} \subseteq \mathcal{D}_{\mathrm{RS}}^{\star}$.
Nevertheless, such restriction to a single power variable turns out to be lossless in the DoF sense.

Next, we observe that the private DoF variables in \eqref{eq:DoF_region_RS_a} can be easily eliminated by
replacing each $d_{i}^{(\mathrm{p})}$ with $d_{i} - d_{i}^{(\mathrm{c})}$, for all $i \in \mathcal{K}$.
After this elimination, the set of inequalities in \eqref{eq:DoF_region_RS_a} are equivalently expressed as
\begin{subequations}
  \label{eq:DoF_RS_region_a}
\begin{align}
  & - d_{i}^{(\mathrm{c})}  \leq 0, \ i \in \mathcal{K} \\
  \label{eq:DoF_RS_region_a_min_1}
  &d_{i} - d_{i}^{(\mathrm{c})}  \leq \alpha_{i}, \ i \in \mathcal{K} \\
  \label{eq:DoF_RS_region_a_min_2}
  &d_{i} - d_{i}^{(\mathrm{c})}  \leq a, \ i \in \mathcal{K} \\
  \label{eq:DoF_RS_region_a_1}
  &\sum_{i \in \mathcal{K}} d_{i}^{(\mathrm{c})} \leq 1 - a \\
  &0 \leq a \leq  1
\end{align}
\end{subequations}
where \eqref{eq:DoF_RS_region_a_min_1} and \eqref{eq:DoF_RS_region_a_min_2} are equivalent to \eqref{eq:DoF_region_RS_a_min}.
The following step is to eliminate the remaining auxiliary variables, i.e. $(\mathbf{d}^{(\mathrm{c})},a)$,
from \eqref{eq:DoF_RS_region_a}.
Observing that the inequalities in \eqref{eq:DoF_RS_region_a} describe a polyhedron in a $(2K+1)$-dimensional space,
elimination is accomplished using a Fourier-Motzkin procedure, together with induction to
address arbitrary $K$.
A detailed explanation of this procedure is presented in the following section.

As seen towards the end of next section, eliminating $(\mathbf{d}^{(\mathrm{c})},a)$ from \eqref{eq:DoF_RS_region_a} yields
a representation of $\mathcal{D}_{\mathrm{RS}}$ which is identical to \eqref{eq:DoF_region_single_subchannel},
giving rise to the following result.
\begin{theorem}
\label{Theorem:single_subchannel}
The achievable DoF region $\mathcal{D}_{\mathrm{RS}}$ described in \eqref{eq:DoF_region_RS_a} coincides with the optimal DoF region $\mathcal{D}$ described in \eqref{eq:DoF_region_single_subchannel}.
\end{theorem}
From Theorem \ref{Theorem:single_subchannel},
it follows that when designing the rate-splitting achievability scheme,
in addition to the flexible common DoF assignment,
\emph{varying only a single power variable} $a$, as opposed to the
$K$ variables $a_{1},\ldots,a_{K}$ used in \cite{Piovano2017}, is \emph{sufficient} to achieve all
points of the DoF region $\mathcal{D}$.
\section{Inductive Fourier-Motzkin Elimination}
In this section, we eliminate the auxiliary variables  $(\mathbf{d}^{(\mathrm{c})},a)$
from \eqref{eq:DoF_RS_region_a}.
First, we impose a further restriction by replacing the inequality in \eqref{eq:DoF_RS_region_a_1} with the equality
$\sum_{i \in \mathcal{K}} d_{i}^{(\mathrm{c})} = 1 - a $. While in principle this restriction yields an achievable DoF
region contained in $\mathcal{D}_{\mathrm{RS}}$, it turns out to be inconsequential.

Next, we eliminate the power control variable $a$ in \eqref{eq:DoF_RS_region_a} by replacing it with
$1 - \sum_{i \in \mathcal{K}} d_{i}^{(\mathrm{c})} = 1 - \mathbf{d}^{(\mathrm{c})}(\mathcal{K})$.
This in turn leaves us with the following set of inequalities
\begin{subequations}
\label{eq:D_m_0}
\begin{align}
  d_{i} - d_{i}^{(\mathrm{c})}  & \leq \alpha_{i}, \ i \in \mathcal{K} \\
   - d_{i}^{(\mathrm{c})}       & \leq 0, \ i \in \mathcal{K} \\
  d_{i} + \mathbf{d}^{(\mathrm{c})}\big( \mathcal{K}\setminus \{i\} \big)  & \leq 1, \ i \in \mathcal{K} \\
  \mathbf{d}^{(\mathrm{c})}( \mathcal{K})  & \leq 1.
\end{align}
\end{subequations}
In what follows, we focus on the set of inequalities in \eqref{eq:D_m_0} and remove the common DoF
variables $\mathbf{d}^{(\mathrm{c})}$ using Fourier-Motzkin elimination \cite[Appendix D]{ElGamal2011}.
This comprises $K$ steps, where in each step $k \in \mathcal{K}$ we eliminate the common DoF variable
$d^{(\mathrm{c})}_{k}$.
We further complement the elimination procedure with mathematical induction so that it applies to any arbitrary $K$.
To gain insight into the induction hypothesis, we start by manually carrying out the first 2 steps of the elimination.
\subsection{Fourier-Motzkin Elimination: Step $1$}
To eliminate $d_{1}^{(\mathrm{c})}$, we first group the set inequalities in \eqref{eq:D_m_0} into the three following categories depending on the presence and sign of $d_{1}^{(\mathrm{c})}$ on the left-hand-side of the inequalities.
\begin{itemize}
\item Inequalities without $d_{1}^{(\mathrm{c})}$:
\begin{subequations}
\label{eq:D_m_1_no}
\begin{align}
  d_{i} - d_{i}^{(\mathrm{c})} &\leq  \alpha_{i}, \ i \in \langle 2:K \rangle \\
   - d_{i}^{(\mathrm{c})}      & \leq  0, \ i \in \langle 2:K \rangle \\
\label{eq:D_m_1_no_3}
  d_{1} + \mathbf{d}^{(\mathrm{c})}\big( \mathcal{K}\setminus \{1\} \big) & \leq 1.
\end{align}
\end{subequations}
\item Inequalities with $-d_{1}^{(\mathrm{c})}$:
\begin{subequations}
\label{eq:D_m_1_N}
\begin{align}
  d_{1} - d_{1}^{(\mathrm{c})} &\leq  \alpha_{1} \\
   - d_{1}^{(\mathrm{c})}      & \leq  0.
\end{align}
\end{subequations}
\item Inequalities with $+d_{1}^{(\mathrm{c})}$:
\begin{subequations}
\label{eq:D_m_1_P}
\begin{align}
  d_{i} + d_{1}^{(\mathrm{c})} + \mathbf{d}^{(\mathrm{c})}\big( \mathcal{K}\setminus \{1,i\} \big) &\leq  1, \ i \in \langle 2:K \rangle \\
   d_{1}^{(\mathrm{c})} + \mathbf{d}^{(\mathrm{c})}\big( \mathcal{K}\setminus \{1\} \big)     & \leq  1.
\end{align}
\end{subequations}
\end{itemize}
Next, we eliminate the variable $d_{1}^{(\mathrm{c})}$ by adding each inequality in \eqref{eq:D_m_1_N} to every inequality in \eqref{eq:D_m_1_P}
(see \cite[Appendix D]{ElGamal2011}).
This procedure yields the following set of inequalities:
\begin{subequations}
\label{eq:D_m_1_N_P}
\begin{align}
d_{1} + d_{i} + \mathbf{d}^{(\mathrm{c})}\big( \mathcal{K}\setminus \{1,i\} \big) &\leq  1 + \alpha_{1}, \ i \in \langle 2:K \rangle \\
d_{i} + \mathbf{d}^{(\mathrm{c})}\big( \mathcal{K}\setminus \{1,i\} \big) &\leq  1 , \ i \in \langle 2:K \rangle \\
\label{eq:D_m_1_N_P_3}
d_{1} + \mathbf{d}^{(\mathrm{c})}\big( \mathcal{K}\setminus \{1\} \big)     & \leq  1 + \alpha_{1} \\
\label{eq:D_m_1_N_P_4}
\mathbf{d}^{(\mathrm{c})}\big( \mathcal{K}\setminus \{1\} \big)     & \leq  1.
\end{align}
\end{subequations}
At this point, we are left with the inequalities in  \eqref{eq:D_m_1_no} and \eqref{eq:D_m_1_N_P}, where
$d_{1}^{(\mathrm{c})}$  has been eliminated.
We observe that \eqref{eq:D_m_1_N_P_3} is redundant as it is implied by \eqref{eq:D_m_1_no_3}.
Moreover, since $d_{1} \geq 0$, the inequality in \eqref{eq:D_m_1_N_P_4} is redundant as it
is also implied by \eqref{eq:D_m_1_no_3}. It follows that at the end of step $1$ (and at the beginning of step $2$), we have
the following set of inequalities
\begin{subequations}
\label{eq:D_m_2}
\begin{align}
  d_{i} - d_{i}^{(\mathrm{c})} &\leq  \alpha_{i}, \ i \in \langle 2:K \rangle \\
   - d_{i}^{(\mathrm{c})}      & \leq  0, \ i \in \langle 2:K \rangle \\
d_{1} + d_{i} + \mathbf{d}^{(\mathrm{c})}\big( \mathcal{K}\setminus \{1,i\} \big) &\leq  1 + \alpha_{1}, \ i \in \langle 2:K \rangle \\
d_{i} + \mathbf{d}^{(\mathrm{c})}\big( \mathcal{K}\setminus \{1,i\} \big) &\leq  1 , \ i \in \langle 2:K \rangle \\
d_{1} + \mathbf{d}^{(\mathrm{c})}\big( \mathcal{K}\setminus \{1\} \big) & \leq 1.
\end{align}
\end{subequations}
\subsection{Fourier-Motzkin Elimination: Step $2$}
For the purpose of eliminating the variable $d_{2}^{(\mathrm{c})}$, we categorize the inequalities in \eqref{eq:D_m_2} as follows:
\begin{itemize}
\item Inequalities without $d_{2}^{(\mathrm{c})}$:
\begin{subequations}
\label{eq:D_m_2_no}
\begin{align}
  d_{i} - d_{i}^{(\mathrm{c})} &\leq  \alpha_{i}, \ i \in \langle 3:K \rangle \\
   - d_{i}^{(\mathrm{c})}      & \leq  0, \ i \in \langle 3:K \rangle \\
\label{eq:D_m_2_no_3}
d_{1} + d_{2} + \mathbf{d}^{(\mathrm{c})}\big( \mathcal{K}\setminus \{1,2\} \big) & \leq 1 + \alpha_{1} \\
d_{2} + \mathbf{d}^{(\mathrm{c})}\big( \mathcal{K}\setminus \{1,2\} \big) & \leq 1.
\end{align}
\end{subequations}
\item Inequalities with $-d_{2}^{(\mathrm{c})}$:
\begin{subequations}
\label{eq:D_m_2_N}
\begin{align}
  d_{2} - d_{2}^{(\mathrm{c})} &\leq  \alpha_{2} \\
   - d_{2}^{(\mathrm{c})}      & \leq  0.
\end{align}
\end{subequations}
\item Inequalities with $+d_{2}^{(\mathrm{c})}$:
\begin{subequations}
\label{eq:D_m_2_P}
\begin{align}
  &d_{1} \! + \! d_{i} \! + \! d_{2}^{(\mathrm{c})} \! + \! \mathbf{d}^{(\mathrm{c})}\big( \mathcal{K}\setminus \{1,2,i\} \big) \leq  1 \! + \! \alpha_{1}
  , i \in \langle 3:K \rangle \\
  &d_{i}  +  d_{2}^{(\mathrm{c})}  +  \mathbf{d}^{(\mathrm{c})}\big( \mathcal{K}\setminus \{1,2,i\} \big) \leq  1, i \in \langle 3:K \rangle \\
  &d_{1} + d_{2}^{(\mathrm{c})} + \mathbf{d}^{(\mathrm{c})}\big( \mathcal{K}\setminus \{1,2\} \big)      \leq  1.
\end{align}
\end{subequations}
\end{itemize}
Now we eliminate $d_{2}^{(\mathrm{c})}$ by adding the inequalities in \eqref{eq:D_m_2_N} and \eqref{eq:D_m_2_P}.
This yields:
\begin{subequations}
\label{eq:D_m_2_N_P}
\begin{align}
  &d_{1} \! + \! d_{2} \! + \! d_{i}  \! +\! \mathbf{d}^{(\mathrm{c})}\big( \! \mathcal{K} \! \setminus \! \{\! 1,2,i \! \} \! \big) \! \leq \!  1 \!+\! \alpha_{1} \!+\! \alpha_{2}
  , i \! \in \! \langle 3\! : \! K \rangle \\
  &
    d_{1} + d_{i} + \mathbf{d}^{(\mathrm{c})}\big( \mathcal{K}\setminus \{1,2,i\} \big) \leq  1 + \alpha_{1}
  , \ i \in \langle 3:K \rangle \\
  &d_{2} + d_{i}  + \mathbf{d}^{(\mathrm{c})}\big( \mathcal{K}\setminus \{1,2,i\} \big) \leq  1 + \alpha_{2}, \ i \in \langle 3:K \rangle \\
  \label{eq:D_m_2_N_P_4}
  &d_{i}  + \mathbf{d}^{(\mathrm{c})}\big( \mathcal{K}\setminus \{1,2,i\} \big) \leq  1, \ i \in \langle 3:K \rangle \\
  \label{eq:D_m_2_N_P_5}
  &d_{1} + d_{2} + \mathbf{d}^{(\mathrm{c})}\big( \mathcal{K}\setminus \{1,2\} \big)      \leq  1 + \alpha_{2} \\
  &d_{1}  + \mathbf{d}^{(\mathrm{c})}\big( \mathcal{K}\setminus \{1,2\} \big)      \leq  1.
\end{align}
\end{subequations}
After eliminating $d_{2}^{(\mathrm{c})}$, we are left with the inequalities in \eqref{eq:D_m_2_no} and
\eqref{eq:D_m_2_N_P}.
Moreover, it can be seen that the inequality in \eqref{eq:D_m_2_no_3} is now redundant as it is implied by the inequality in
\eqref{eq:D_m_2_N_P_5}.
The remaining inequalities in \eqref{eq:D_m_2_no} and
\eqref{eq:D_m_2_N_P} are expressed in compact form as follows:
\begin{subequations}
\label{eq:D_m_3}
\begin{align}
&d_{i} - d_{i}^{(\mathrm{c})} \leq  \alpha_{i}, \ i \in \langle 3:K \rangle \\
&   - d_{i}^{(\mathrm{c})}       \leq  0, \ i \in \langle 3:K \rangle \\
\nonumber
& \mathbf{d}(\mathcal{S}) + d_{i}  + \mathbf{d}^{(\mathrm{c})}\big( \mathcal{K}\setminus \{1,2,i\} \big) \leq  1 + \bm{\alpha}(\mathcal{S}), \\
\label{eq:D_m_3_3}
& \quad \quad \mathcal{S} \subseteq \{1,2\}, i \in \langle 3:K \rangle \\
\label{eq:D_m_3_4}
& \mathbf{d}(\mathcal{S})  \!  +  \! \mathbf{d}^{(\mathrm{c})}\big( \! \mathcal{K} \! \! \setminus \! \! \{1,2\} \! \big) \! \leq \! 1 \!+\! \bm{\alpha}\big(\! \mathcal{S} \! \setminus \! \{ \min \mathcal{S} \} \! \big), \mathcal{S} \! \subseteq \! \{ \!1,2 \! \}.
\end{align}
\end{subequations}
In the above, we use the convention that  $\mathcal{S} = \emptyset$ is a subset of $\{1,2\}$
so that the inequalities in \eqref{eq:D_m_2_N_P_4} are included in \eqref{eq:D_m_3_3}.
On the other hand, by setting $\mathcal{S} = \emptyset$ in \eqref{eq:D_m_3_4}, we obtain the inequality
$\mathbf{d}^{(\mathrm{c})}\big( \mathcal{K}\setminus \{1,2\} \big) \leq  1 $, which is implied by \eqref{eq:D_m_2_N_P_5}
and hence has no influence.
\subsection{Fourier-Motzkin Elimination: Step $k + 1$}
Guided by the first two elimination steps, we now construct the induction hypothesis.
Suppose that after $k$ steps of the procedure, where $k \in \langle 1 : K-2 \rangle$, the variables $d_{1}^{(\mathrm{c})},\ldots,d_{k}^{(\mathrm{c})}$
are eliminated and we are left with the set of inequalities:
\begin{subequations}
\label{eq:D_end_of_m}
\begin{align}
&  d_{i} - d_{i}^{(\mathrm{c})} \leq  \alpha_{i}, \ i \in \langle k+1:K \rangle \\
&   - d_{i}^{(\mathrm{c})}       \leq  0, \ i \in \langle k+1:K \rangle \\
\nonumber
& \mathbf{d}(\mathcal{S}) + d_{i}  + \mathbf{d}^{(\mathrm{c})}\big( \mathcal{K}\setminus \big\{ \{i\}\cup\langle1:k \rangle \big\} \big) \leq  1 + \bm{\alpha}(\mathcal{S}),  \\
&
\quad \quad  \mathcal{S} \subseteq \langle 1:k \rangle, i \in \langle k+1:K \rangle \\
& \mathbf{d}( \! \mathcal{S} \!) \!  + \! \mathbf{d}^{(\mathrm{c})}\big( \! \mathcal{K} \! \! \setminus \! \! \langle 1:k \rangle \! \big) \! \leq \! 1 \! + \! \bm{\alpha}\big( \! \mathcal{S} \!  \! \setminus \!  \! \{\min
\mathcal{S} \} \! \big), \mathcal{S} \! \subseteq \! \langle 1:k \rangle.
\end{align}
\end{subequations}
Note that the above hypothesis is consistent with the results from steps $1$ and $2$.
Next, we show that by the end of step $k+1$, the variable $d_{k+1}^{(\mathrm{c})}$ is eliminated and
we obtain a set of inequalities similar to \eqref{eq:D_end_of_m}, except that $k$ in \eqref{eq:D_end_of_m}  is replaced with $k+1$.
For this purpose, we group the inequalities in \eqref{eq:D_end_of_m} into the following three categories:
\begin{itemize}
\item Inequalities without $d_{k+1}^{(\mathrm{c})}$:
\begin{subequations}
\label{eq:D_m_plus_1_no}
\begin{align}
 & d_{i} - d_{i}^{(\mathrm{c})} \leq  \alpha_{i}, \ i \in \langle k+2:K \rangle \\
 &  - d_{i}^{(\mathrm{c})}       \leq  0, \ i \in \langle k+2:K \rangle \\
\nonumber
& \mathbf{d}(\mathcal{S}) + d_{k+1} + \mathbf{d}^{(\mathrm{c})}\big( \mathcal{K}\setminus \langle 1:k+1 \rangle \big)  \leq 1 + \bm{\alpha}(\mathcal{S}), \\
\label{eq:D_m_plus_1_no_3}
& \quad \quad   \mathcal{S} \subseteq \langle 1:k \rangle.
\end{align}
\end{subequations}
\item Inequalities with $-d_{k+1}^{(\mathrm{c})}$:
\begin{subequations}
\label{eq:D_m_plus_1_N}
\begin{align}
  d_{k+1} - d_{k+1}^{(\mathrm{c})} &\leq  \alpha_{k+1} \\
   - d_{k+1}^{(\mathrm{c})}      & \leq  0.
\end{align}
\end{subequations}
\item Inequalities with $+d_{k+1}^{(\mathrm{c})}$:
\begin{subequations}
\label{eq:D_m_plus_1_P}
\begin{align}
\nonumber
&\mathbf{d}(\mathcal{S}) + d_{i}  + d_{k+1}^{(\mathrm{c})} + \mathbf{d}^{(\mathrm{c})}\big( \mathcal{K} \! \setminus \! \big\{ \{i\}\cup \langle 1:k+1 \rangle \big\} \big) \leq \\
& \quad \quad  1 \! + \! \bm{\alpha}(\mathcal{S}),
\mathcal{S} \! \subseteq \! \langle 1:k \rangle, i \! \in \! \langle  k+2 \! : \! K  \rangle \\
\nonumber
& \mathbf{d}(\mathcal{S}) + d_{k+1}^{(\mathrm{c})}  + \mathbf{d}^{(\mathrm{c})}\big( \mathcal{K}\setminus \langle 1:k+1 \rangle \big) \leq  \\
& \quad \quad 1 + \bm{\alpha}\big(\mathcal{S} \setminus \{\min
\mathcal{S} \} \big), \ \mathcal{S} \subseteq \langle 1:k \rangle.
\end{align}
\end{subequations}
\end{itemize}
Now we eliminate $d_{k+1}^{(\mathrm{c})}$ by adding the inequalities in \eqref{eq:D_m_plus_1_N} and \eqref{eq:D_m_plus_1_P},
from which we obtain
\begin{subequations}
\label{eq:D_m_plus_1_N_P}
\begin{align}
\nonumber
& \mathbf{d}(\mathcal{S}) \! + \! d_{k+1} \! + \! d_{i} \!  + \! \mathbf{d}^{(\mathrm{c})}\big( \mathcal{K} \! \setminus \! \big\{ \{i\}\cup \langle 1:k+1 \rangle \big\} \big) \! \leq \\
\label{eq:D_m_plus_1_N_P_1}
&
1 \! + \! \bm{\alpha}(\mathcal{S} \! \cup \! \{k+1\}),
\mathcal{S} \! \subseteq \! \langle 1\!:\!k  \rangle, i \! \in \! \langle  k\!+\!2 \! : \! K  \rangle \\
\nonumber
&\mathbf{d}(\mathcal{S}) + d_{i}  + \mathbf{d}^{(\mathrm{c})}\big( \mathcal{K}  \setminus  \big\{ \{i\}\cup \langle 1:k+1 \rangle \big\} \big)  \leq   1  + \bm{\alpha}(\mathcal{S}), \\
\label{eq:D_m_plus_1_N_P_2}
& \quad \quad  \mathcal{S}  \subseteq  \langle 1:k \rangle, i \in  \langle  k+2 :  K  \rangle \\
\nonumber
& \mathbf{d}(\mathcal{S}) + d_{k+1}  + \mathbf{d}^{(\mathrm{c})}\big( \mathcal{K}\setminus \langle 1:k+1 \rangle  \big) \leq  \\
\label{eq:D_m_plus_1_N_P_3}
& 1 + \bm{\alpha}\big(\mathcal{S} \cup  \{k+1\} \setminus \{\min
\mathcal{S} \} \big),  \mathcal{S} \subseteq \langle 1:k \rangle \\
\nonumber
&\mathbf{d}(\mathcal{S})  + \mathbf{d}^{(\mathrm{c})}\big( \mathcal{K}\setminus \langle 1:k+1 \rangle  \big) \leq  1 + \bm{\alpha}\big(\mathcal{S} \setminus \{\min
\mathcal{S} \} \big), \\
\label{eq:D_m_plus_1_N_P_4}
& \quad \quad  \mathcal{S} \subseteq \langle 1:k \rangle.
\end{align}
\end{subequations}
After the elimination, we are left with the inequalities in \eqref{eq:D_m_plus_1_no} and \eqref{eq:D_m_plus_1_N_P}.
Next, we observe that for any $\mathcal{S} \subseteq \langle 1:k \rangle$, we have $k+1 > j$ (and hence $\alpha_{k+1} \leq \alpha_{j}$) for all $j \in \mathcal{S}$, and hence
\begin{multline}
\label{eq:m_plus_1_redundant}
\bm{\alpha}\big(\mathcal{S} \cup  \{k+1\} \setminus \{\min
\mathcal{S} \} \big) = \bm{\alpha}(\mathcal{S}) + \alpha_{k+1} - \max_{j \in \mathcal{S}} \alpha_{j}  \\
\leq \bm{\alpha}(\mathcal{S}), \ \forall \mathcal{S} \subseteq \langle 1:k \rangle.
\end{multline}
From \eqref{eq:m_plus_1_redundant}, we conclude that the inequalities in \eqref{eq:D_m_plus_1_no_3} are redundant as they are implied by the ones in \eqref{eq:D_m_plus_1_N_P_3}.
It follows that at the end of step $k+1$, the variable $d_{k+1}^{(\mathrm{c})}$ is eliminated and
we are left with the set of inequalities given by:
\begin{subequations}
\label{eq:D_end_of_m_plus_1}
\begin{align}
& d_{i} - d_{i}^{(\mathrm{c})} \leq  \alpha_{i}, \ i \in \langle k+2:K \rangle \\
& - d_{i}^{(\mathrm{c})}      \leq  0, \ i \in \langle k+2:K \rangle \\
\nonumber
& \mathbf{d}(\mathcal{S}) + d_{i}  + \mathbf{d}^{(\mathrm{c})}\big( \mathcal{K}\setminus \big\{ \{i\}\cup \langle 1:k+1 \rangle \big\} \big) \leq  1 + \bm{\alpha}(\mathcal{S}), \\
\label{eq:D_end_of_m_plus_1_3}
& \quad \quad  \mathcal{S} \subseteq \langle 1:k+1 \rangle, i \in \langle k+2:K \rangle \\
\nonumber
& \mathbf{d}(\mathcal{S})   + \mathbf{d}^{(\mathrm{c})}\big( \mathcal{K}\setminus \langle 1:k+1 \rangle \big) \leq  1 + \bm{\alpha}\big(\mathcal{S} \setminus \{\min
\mathcal{S} \} \big), \\
\label{eq:D_end_of_m_plus_1_4}
& \quad \quad  \mathcal{S} \subseteq \langle 1:k+1 \rangle.
\end{align}
\end{subequations}
Note that \eqref{eq:D_end_of_m_plus_1_3} corresponds to \eqref{eq:D_m_plus_1_N_P_1} and \eqref{eq:D_m_plus_1_N_P_2},
while \eqref{eq:D_end_of_m_plus_1_4} corresponds to \eqref{eq:D_m_plus_1_N_P_3} and \eqref{eq:D_m_plus_1_N_P_4}.
It is evident that the set of inequalities in \eqref{eq:D_end_of_m_plus_1} take the same form of the set of inequalities in \eqref{eq:D_end_of_m},
with the difference that $k+1$ replaces $k$.
\subsection{Fourier-Motzkin Elimination: Step $K$}
From the above induction hypothesis, by setting $k = K-2$, it can be seen that at the end of step $k+1 = K-1$ of the procedure,
we obtain the following set of inequalities:
\begin{subequations}
\label{eq:D_start_of_K}
\begin{align}
  d_{K} - d_{K}^{(\mathrm{c})} &\leq  \alpha_{K} \\
   - d_{K}^{(\mathrm{c})}      & \leq  0 \\
\mathbf{d}(\mathcal{S}) + d_{K}   &\leq  1 + \bm{\alpha}(\mathcal{S}), \
\mathcal{S} \subseteq \langle 1:K-1 \rangle \\
\mathbf{d}(\mathcal{S})  \! + \! d^{(\mathrm{c})}_{K} &\leq  1 \! + \! \bm{\alpha}\big( \! \mathcal{S} \! \setminus \! \{\min
\mathcal{S} \} \! \big), \mathcal{S} \subseteq \langle 1:K-1 \rangle.
\end{align}
\end{subequations}
Therefore, after eliminating $d_{K}^{(\mathrm{c})}$ in step $K$, we are left with the following set of inequalities:
\begin{subequations}
\label{eq:D_end_of_K}
\begin{align}
\label{eq:D_end_of_K_1}
\mathbf{d}\big(\mathcal{S}' \cup \{K\} \big)  &\leq  1 + \bm{\alpha}(\mathcal{S}'), \
\mathcal{S}' \subseteq \langle 1:K-1 \rangle \\
\label{eq:D_end_of_K_2}
\mathbf{d}(\mathcal{S})   &\leq  1 + \bm{\alpha}\big(\mathcal{S} \setminus \{\min
\mathcal{S} \} \big), \ \mathcal{S} \subseteq \mathcal{K}.
\end{align}
\end{subequations}
Finally, we show that the set of inequalities in \eqref{eq:D_end_of_K_1} are redundant.
For $\mathcal{S}' = \emptyset$ in \eqref{eq:D_end_of_K_1}, it can be seen that the resulting inequality is included in \eqref{eq:D_end_of_K_2}.
Therefore, we consider a non-empty subset $\mathcal{S}' \subseteq \langle 1:K-1 \rangle$ in \eqref{eq:D_end_of_K_1}
and choose $\mathcal{S} = \mathcal{S}' \cup \{K\}$ in \eqref{eq:D_end_of_K_2} to obtain the corresponding inequality.
Since $K > j$ (and hence $\alpha_{K} \leq \alpha_{j}$) for all $j \in \mathcal{S}'$, we have
\begin{multline}
\label{eq:K_redundant}
\bm{\alpha}\big(\mathcal{S}' \cup \{K\} \setminus \{\min \{
\mathcal{S}',K\} \} \big)  = \bm{\alpha}\big(\mathcal{S}' \cup \{K\} \setminus \{\min \mathcal{S}' \} \big) \\
= \bm{\alpha}(\mathcal{S}') + \alpha_{K} - \max_{j\in \mathcal{S}'}\alpha_{j}
\leq \bm{\alpha}(\mathcal{S}').
\end{multline}
Hence, we conclude that the inequalities in \eqref{eq:D_end_of_K_1} are looser in general compared to the corresponding inequalities in \eqref{eq:D_end_of_K_2}.
This leaves us with \eqref{eq:D_end_of_K_2} in addition to the implicit non-negativity condition $\mathbf{d} \in \mathbb{R}_{+}^{K}$.
Therefore, $\mathcal{D}_{\mathrm{RS}}$ in \eqref{eq:DoF_region_RS_a} is equivalent to $\mathcal{D}$ in \eqref{eq:DoF_region_single_subchannel},
hence proving Theorem \ref{Theorem:single_subchannel}.
\section{Discussion}
\label{sec:Discussion}
As seen from Section \ref{sec:preliminaries}, the proof in \cite{Piovano2017} is of the constructive type, i.e. an
explicit scheme is constructed to achieve each point (or face) in the known DoF region outer bound.
On the other hand, the proof presented in this paper is of the existence type, i.e.
we essentially show that there exists a scheme, specified by tuning some design variables, that achieves
each point in the DoF region outer bound without explicitly optimizing these variables.
Both approaches have their merits and drawbacks.
The former provides insights into the specific tuning of design variables, yet relies on a very
specific exhaustive procedure which is not easy to extend beyond the considered setup.
On the other hand, the latter sacrifices explicit constructions for the sake of a more malleable and general approach
which can potentially lend itself to solve problems beyond the considered setup.
\bibliographystyle{IEEEtran}
\bibliography{References}

\begin{thebibliography}{10}
\providecommand{\url}[1]{#1}
\csname url@samestyle\endcsname
\providecommand{\newblock}{\relax}
\providecommand{\bibinfo}[2]{#2}
\providecommand{\BIBentrySTDinterwordspacing}{\spaceskip=0pt\relax}
\providecommand{\BIBentryALTinterwordstretchfactor}{4}
\providecommand{\BIBentryALTinterwordspacing}{\spaceskip=\fontdimen2\font plus
\BIBentryALTinterwordstretchfactor\fontdimen3\font minus
  \fontdimen4\font\relax}
\providecommand{\BIBforeignlanguage}[2]{{%
\expandafter\ifx\csname l@#1\endcsname\relax
\typeout{** WARNING: IEEEtran.bst: No hyphenation pattern has been}%
\typeout{** loaded for the language `#1'. Using the pattern for}%
\typeout{** the default language instead.}%
\else
\language=\csname l@#1\endcsname
\fi
#2}}
\providecommand{\BIBdecl}{\relax}
\BIBdecl

\bibitem{Vaze2012}
C.~S. {Vaze} and M.~K. {Varanasi}, ``The degree-of-freedom regions of {MIMO}
  broadcast, interference, and cognitive radio channels with no {CSIT},''
  \emph{IEEE Trans. Inf. Theory}, vol.~58, no.~8, pp. 5354--5374, 2012.

\bibitem{Yang2013}
S.~Yang, M.~Kobayashi, D.~Gesbert, and X.~Yi, ``Degrees of freedom of time
  correlated {MISO} broadcast channel with delayed {CSIT},'' \emph{IEEE Trans.
  Inf. Theory}, vol.~59, no.~1, pp. 315--328, 2013.

\bibitem{Tandon2013}
R.~Tandon, S.~A. Jafar, S.~Shamai, and H.~V. Poor, ``On the synergistic
  benefits of alternating {CSIT} for the {MISO} broadcast channel,'' \emph{IEEE
  Trans. Inf. Theory}, vol.~59, no.~7, pp. 4106--4128, 2013.

\bibitem{Rassouli2016}
B.~Rassouli, C.~Hao, and B.~Clerckx, ``{DoF} analysis of the {MIMO} broadcast
  channel with alternating/hybrid {CSIT},'' \emph{IEEE Trans. Inf. Theory},
  vol.~62, no.~3, pp. 1312--1325, 2016.

\bibitem{Davoodi2016}
A.~G. Davoodi and S.~A. Jafar, ``Aligned image sets under channel uncertainty:
  Settling conjectures on the collapse of degrees of freedom under finite
  precision {CSIT},'' \emph{IEEE Trans. Inf. Theory}, vol.~62, no.~10, pp.
  5603--5618, 2016.

\bibitem{Clerckx2016}
B.~Clerckx, H.~Joudeh, C.~Hao, M.~Dai, and B.~Rassouli, ``{Rate splitting for
  MIMO wireless networks: a promising PHY-layer strategy for LTE evolution},''
  \emph{IEEE Commun. Magazine}, vol.~54, no.~5, pp. 98--105, 2016.

\bibitem{Lapidoth2005}
A.~Lapidoth, S.~Shamai, and M.~Wigger, ``On the capacity of fading {MIMO}
  broadcast channels with imperfect transmitter side-information,'' \emph{Proc.
  43rd Annu. Allerton Conf. Commun., Control Comput.}, 2005.

\bibitem{Joudeh2016}
H.~Joudeh and B.~Clerckx, ``Sum-rate maximization for linearly precoded
  downlink multiuser {MISO} systems with partial {CSIT}: A rate-splitting
  approach,'' \emph{IEEE Trans. Commun.}, vol.~64, no.~11, pp. 4847--4861,
  2016.

\bibitem{Joudeh2016a}
H.~Joudeh and B.~Clerckx, ``Robust transmission in downlink multiuser {MISO}
  systems: A rate-splitting approach,'' \emph{IEEE Trans. Signal Process.},
  vol.~64, no.~23, pp. 6227--6242, 2016.

\bibitem{Piovano2017}
E.~Piovano and B.~Clerckx, ``Optimal {DoF} region of the {$K$}-user {MISO BC}
  with partial {CSIT},'' \emph{IEEE Commun. Lett.}, vol.~21, no.~11, pp.
  2368--2371, 2017.

\bibitem{ElGamal2011}
A.~El~Gamal and Y.-H. Kim, \emph{Network information theory}.\hskip 1em plus
  0.5em minus 0.4em\relax Cambridge university press, 2011.

\end{thebibliography}
\end{document}